\documentclass[oldversion,letter]{aa}
\usepackage{epsfig}
\usepackage{natbib}
\usepackage{lscape}
\usepackage{amssymb}
\usepackage{epstopdf}
\usepackage{longtable}
\usepackage{color}
\usepackage{fancyheadings}
\bibpunct{(}{)}{;}{a}{}{,}

\begin{document}
\title{Polar stellar-spots and grazing planetary transits}
\subtitle{possible explanation for the low number of discovered grazing planets}

\author{M. Oshagh\inst{1},\,N. C. Santos\inst{1,2},\,P. Figueira \inst{1}, V. Zh. Adibekyan\inst{1},  A. Santerne\inst{1}, S. C. C. Barros\inst{1}, J. J. G. Lima\inst{1,2}}

\institute{
Instituto de Astrof\' isica e Ci\^encias do Espa\c{c}o, Universidade do Porto, CAUP, Rua das Estrelas, PT4150-762 Porto, Portugal \\
email: {\tt moshagh@astro.up.pt}
\and
Departamento de F{\'i}sica e Astronomia, Faculdade de Ci{\^e}ncias, Universidade do
Porto,Rua do Campo Alegre, 4169-007 Porto, Portugal
}

\date{Received XXX; accepted XXX}

\abstract {We assess a physically feasible explanation for the low number of discovered (near-)grazing planetary transits through all ground and space based transit surveys.
 We performed simulations to generate the synthetic distribution of detectable planets based on their impact parameter, and found that a larger number of (near-)grazing planets should have been detected than have been detected. Our explanation for the insufficient number of (near-)grazing planets is based on a simple assumption that a large number of (near-)grazing planets transit host stars which harbor dark giant polar spot, and thus the transit light-curve vanishes due to the occultation of grazing planet and the polar spot. We conclude by evaluating the properties required of polar spots in order to make disappear the grazing transit light-curve, and we conclude that their properties are compatible with the expected properties from observations.}



\authorrunning{M. Oshagh et al.}
\maketitle
\section{Introduction}
The impact parameter ($b$) of a transiting planet on a circular orbit is defined as $ a \cos(i)/ R_{\star}$, in which $a$ is the orbital semi-major axis, $i$ is the orbital inclination, and $R_{\star}$ is the stellar radius. If $(1-R_{planet}/R_{\star}) < b \leq (1+R_{planet}/R_{\star})$, where $R_{planet}$ is the planet radius, the planet does
not fully cover the stellar disk during its transit, and therefore the planetary transit is said to be grazing. So far, only a handful of (near-)grazing exoplanets, have been detected and confirmed through all ground and space based transit surveys. To be more accurate, only eight such systems have been discovered, namely, WASP-34b \citep{Smalley-11}, WASP-67b \citep{Hellier-12, Mancini-14}, HAT-P-
27/WASP-40 \citep{Beky-11, Anderson-11},  WASP-45b \citep{Anderson-12}, CoRoT-25b \citep{Almenara-13},  Kepler-434b \citep{Almenara-15}, Kepler-447b \citep{Lillo-Box-15}, and CoRoT-33b\citep{Csizmadia-15}. The (near-)grazing planets are interesting targets in the sense that they can be used to detect the gravitational perturbation of small bodies in the systems, such as exomoons and Trojans \citep{Lillo-Box-15}, thus they can provide us important information on planetary formation and evolution. To date, there has been no study to explore the possible causes of low number of discovered (near-)grazing planets.

Large, cool (dark), and long lived stellar spots located near the stellar rotational axis (either on the high-latitude or covering the pole) are common features on stars independent of the stellar rotational velocity or spectral type \citep[e.g.][]{Strassmeier-91, Schuessler-92, Piskunov-94, Strassmeier-98,  Hatzes-98, Strassmeier-02, Jeffers-02, Berdyugina-05}. Observations have revealed that polar spots can reach a filling factor ($f$)\footnote{The filling factor is defined as $f = (R_{spot}/R_{\star})^{2}$, where $R_{spot}$ is the spot radius.} of up to 50\%, and also have lifetime around a decade. These two properties suggest that the polar spots might be formed by a different physical mechanism than the low-to-mid latitude spots \citep{Holzwarth-06, Strassmeier-09, Berdyugina-05}. Different techniques have been used to rule out different biases in various techniques, and therefore, independently confirm and characterize polar spots  \citep{Unruh-95, Bruls-98, Berdyugina-02, Sanchis-Ojeda-13}. Moreover, several theoretical and numerical studies have been carried out to define the proper mechanism which is responsible for the polar spot's formation and persistence \citep[e.g.][]{Schrijver-01, Holzwarth-06, Brown-10, Yadav-15}

It has been shown in several studies that the occultation of spots by a transiting planet can generate anomalies in the transit light-curve, and may lead to wrong estimation of the planetary parameters  \citep[e.g.][]{Sanchis-Ojeda-11b, Oshagh-13b, Sanchis-Ojeda-13, Barros-13}. In this work we first examine if the number of (near-)grazing planets detected so far are lower than physically predicted. Thus, we assess the possibility of explaining the low number of detected (near-)grazing planets, based on a assumption that a large number of (near-)grazing planets occult with the large polar spots during their transit. If a (near-)grazing planet crosses a large polar spot, then the transit depth decreases significantly (considering that the limb-darkening also causes a decrease in the transit depth when compared to a central transit), and this leads to a lower signal-to-noise ratio and yielding that the transit signal can be below the detection threshold and be completely missed (see Figure 1 for a schematic illustration of the our basic assumption). In Sect.2, we describe our
simulation that allows us to compare the number of expected grazing planets with the discovered ones. In Sect.3, we interpret the results obtained in Sect.2 to evaluate our hypothesis of missing grazing planets due to occulation with the polar spots. In Sect.4 we conclude our results.

\begin{figure}
\includegraphics[width=87mm, height=65mm]{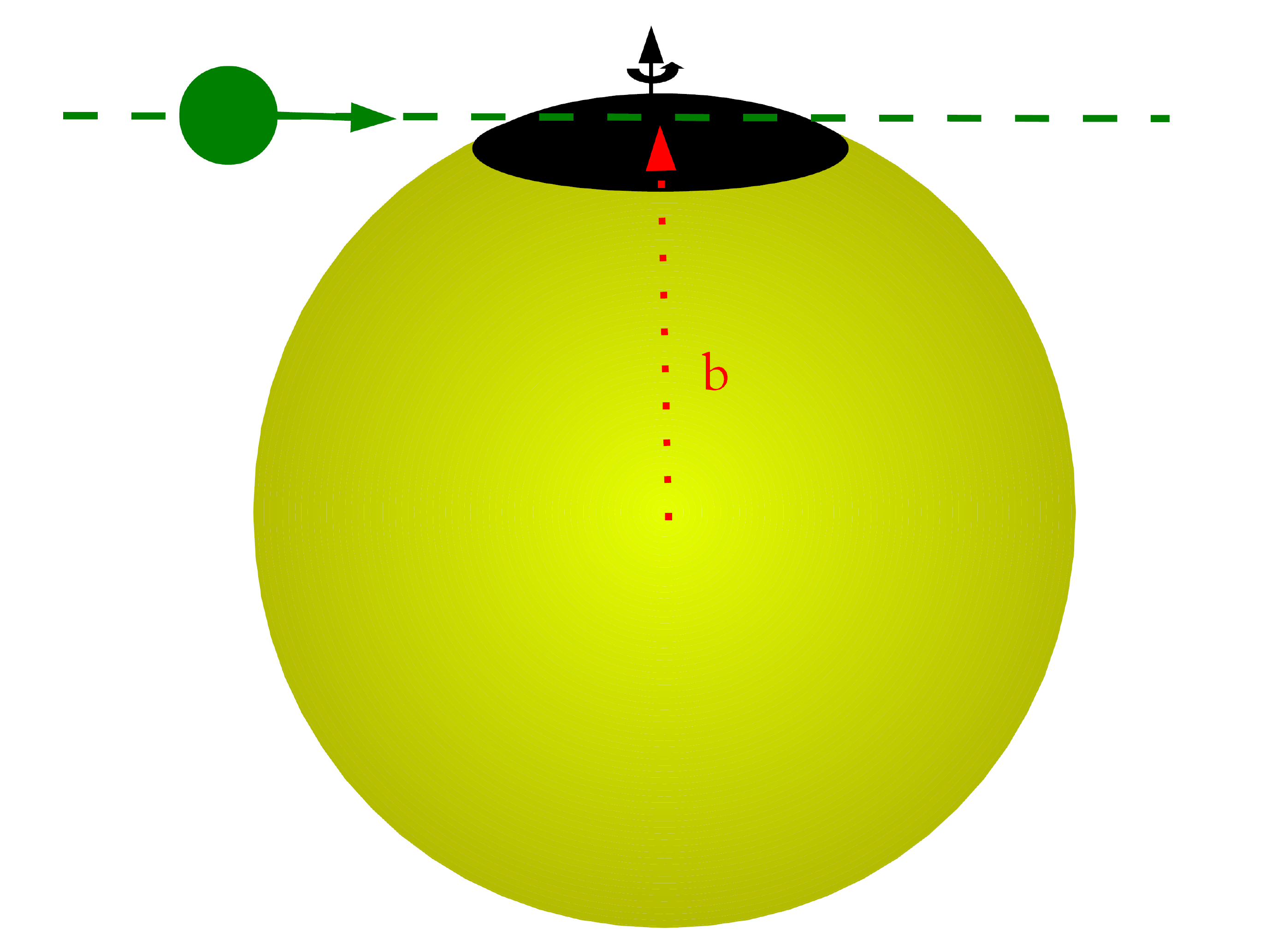}
 \caption{The schematic view of a grazing planet occulting a polar spot.}
  \label{sample-figure}
\end{figure}

\section{Simulation}
In this section we aim to generate a synthetic distribution of transiting planets' impact parameter. To do that we perform
a Monte Carlo simulation by simulating 50 000 transiting planets. The planet radius $(R_{planet})$, orbital period ($P$), semi-major axis ($a$), and stellar radius ($R_{\star}$) were drawn from the observed distributions of all confirmed transiting planets\footnote{The data were obtained through exoplanets.org}. We would like to note that since the observation distribution suffers from observational biases, and it is our 
objective to reproduce the observations even with these (uncharacterized) biases, we 
chose to draw the parameter(s) for the distribution of observed parameter(s). The orbital inclination ($i$) was drawn from a linear distribution in $\cos i$ in the range of 75-90 degree. Note that the range of 75-90 degree was chosen based on the fact that for the inclination smaller than 75 degree the transit probability goes to zero based on the orbital distribution used in this study. The orbital eccentricity and misalignment angle of planets were fixed to zero. By having the semi-major axis, stellar radius, and inclination the impact parameter can be calculated through $ a \cos(i)/ R_{\star}$. The stellar limb darkening coefficients were fixed to the value of the Sun $u1=0.29 $ and $u2=0.34$ \citep{Claret-11}. The stellar inclination was fixed to 90 degree, which means it is seen edge-on. We also consider that the star does not have any spots.

We assign the transiting planet and host star's parameters of each system in the SOAP-T tool\footnote{The SOAP-T tool can generate the light-curves and radial velocity variations for systems consisting of a rotating spotted star with a transiting planet. More details about the tool can be obtained in \citep{Oshagh-13a} or http://www.astro.up.pt/resources/soap-t/}. We would like to mention that SOAP-T requires both the orbital period and semi-major axis thus does not require the stellar mass. Then SOAP-T generates the transit light-curve of the each system. If the transit light-curve shows a transit depth larger than 225 ppm (which means transit depth larger than $3\sigma$, when we consider the $\sigma$ around 75 ppm due to the granulation noise for solar type stars \citep{Gilliland-11}), and also has a duration longer than an hour (in other words the transit light-curve at least shows more than or equal to three points in the transit by assuming the \textit{Kepler}'s long cadence 30 minuets), thus we consider the transiting planet as a ``\textit{detectable}", and if not we consider it as a ``\textit{not-detected}". As a consequence, we can depict the impact parameter distribution of \textit{detectable} planets. 

We would like to note that the granulation noise level strongly depends on the spectral type of host star \citep{Gilliland-11}, therefore, for different stars the detection limit on transit depth can be smaller than the threshold value used (225 ppm). Furthermore, the transit duration limit that we considered (at least three points in one transit) can be a strict limit, because in reality by phase-folding several transits, and thus increasing the signal-to-noise ratio, observers are able to detect planets even with a shorter transit duration than this limit. Thus the proper transit duration limit should be defined as $3\sigma/\sqrt{n}$, which $n$ is the number of observed transits. However, since we are interested in the minimum number of 
planets to be detected, then by assuming these strict limits 
(on the transit depth and also duration) we shall have a 
conservative estimate of the minimum number of grazing planets. As
we show in Section 3, this expected number is still more than that of  observed. As we noticed, planets can be detected well below the 225 ppm 
transit depth limit by observing more transits, down to 99 ppm (e.g. Kepler-90b \citep{Cabrera-14}) or down to $\sim$ 20 ppm (e.g. Kepler-37b \citep{Barclay-13}). This means that the discrepancy between observed and expected numbers of grazing transits can be even higher than we present in Section 3, but we do not study the issue in details. We also note that our simulation is based on some simplistic assumptions, such as fixing the star's limb darkening coefficient, stellar inclination, and orbital eccentricity. However, it still can deliver some insightful and quite realistic approximation about the estimated number of grazing planets. We would like to note that in the presence of polar spots, then limb darkening also changes (because it is temperature dependent), which can affect the transit depth as a second-order effect. Exploring this effect is beyond the scope of this paper.

Comparing the distribution of impact parameter of the known transit exoplanets and the synthetic ones, delivered by our simulation, can provide a informative insight about the significance of missing (near-)grazing planets. 

\section{Results and Interpretation}
Figure 2 presents the synthetic impact parameter's distribution obtained from our simulation in Sect.2 in green histogram, and the confirmed transiting planets' impact parameter distribution in blue histogram. As the results show, in the roughly estimated region of (near-)grazing planets (region between the dashed and dotes lines in the Figure 2.), we would expect to detect more (near-)grazing planets compare to what have been detected. In order to quantify the significance of discrepancies between the simulation and confirmed transiting planets, in the region of (near-)grazing ($0.9 \leq b \leq 1.1$), we performed the two sample Kolmogorov-Smirnoff (K-S) and the  Anderson–Darling
(A-D)\footnote{A-D test is similar to K-S , but more sensitive because it give  more weight to the tails of distribution
 (https://asaip.psu.edu/Articles/beware-the-kolmogorov-smirnov-test). } statistical tests. The K-S and A-D statistics yield a $P_{KS}=0.005$, and $P_{AD}=0.00031$, respectively. These results present strong 
evidence that two samples should come from completely different underlying distribution. The K-S and A-D tests for ($b \leq 0.9$) deliver $P_{KS}=0.30$, and $P_{AD}=0.22$, respectively, which suggest that we cannot reject the hypothesis that they 
come from the same distribution. Therefore, this disagreement supports our hypothesis that there should be more grazing planets which their signal have been vanished due to occultation with the giant polar spots. The red histogram in Figure 2 displays the \textit{Kepler}'s planet candidates, known as the \textit{Kepler} Object of Interest (KOI), impact parameter's distribution, which shows large number grazing planets candidate, however, they have to be confirmed. As a speculative interpretation from comparing \textit{Kepler}'s planet candidates (KOIs) and synthetic distribution one can 
predict that at least some of the grazing planet candidates should be real, and if 
confirmed they will partially fill the observed ``gap" of (near-)grazing transiting planets.

\begin{figure}
\includegraphics[width=95mm, height=70mm, trim=10mm 0 0 0]{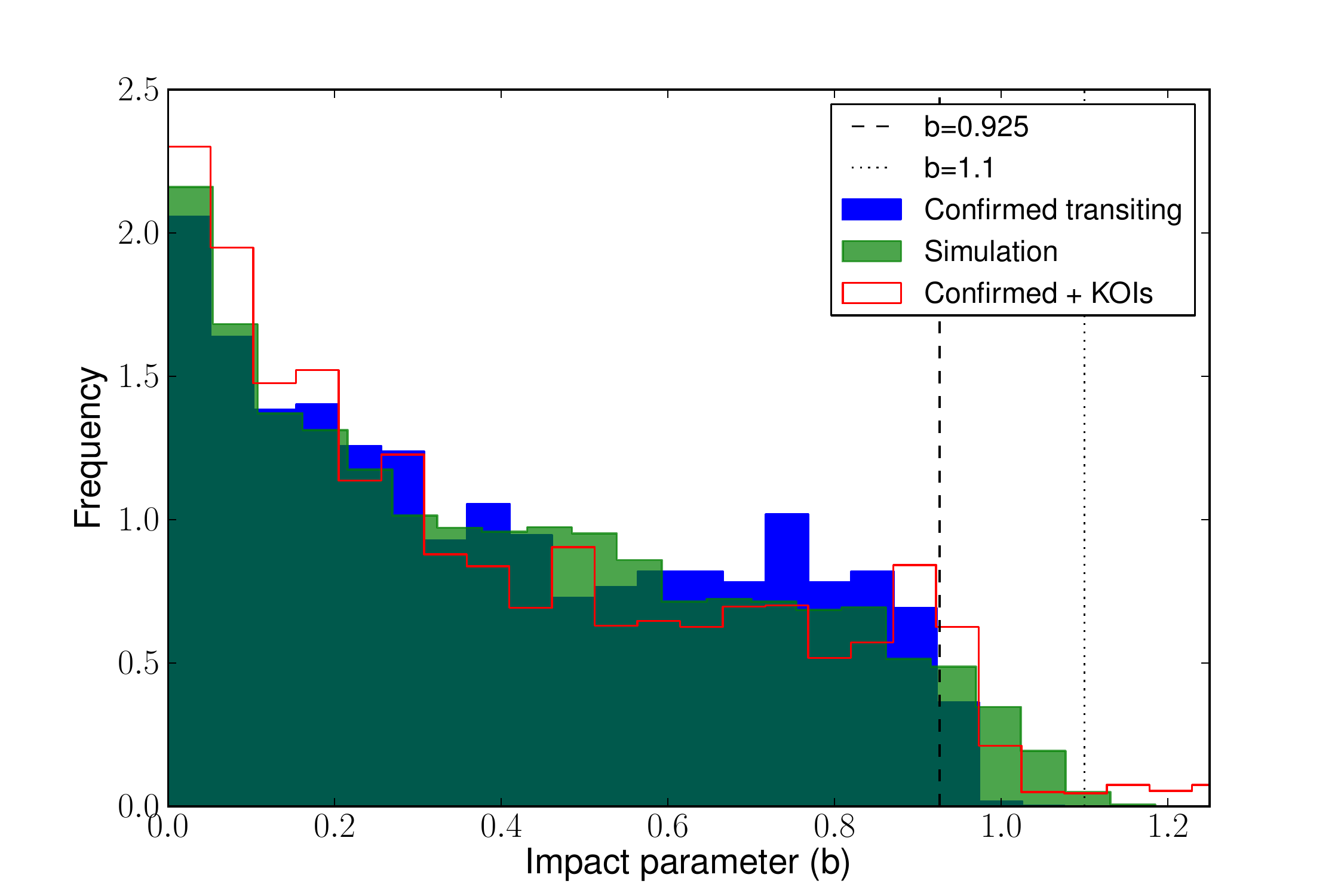}
 \caption{The impact parameter distribution of all confirmed transiting planets (blue histogram). The synthetic impact parameter distribution yielded from our simulation (green histogram). The impact parameter distribution of \textit{Kepler}'s planet candidates (KOI)(red histogram).}
  \label{sample-figure}
\end{figure}

In order to probe if the polar spot's property required to make disappear a grazing transit light-curve has physically feasible values or not, we performed a simple test. We again used the SOAP-T tool \citep{Oshagh-13a} to generate transit light-curves of four (near-)grazing transiting planets, with $b=0.90, 0.95, 1.00,$ and $1.05$. The transiting planets were Jupiter-size planets with radius of $R_{planet}=0.1 R_{\star}$, on a 3-day orbit with semi-major axis of $a= 10 R_{\odot}$. The stellar radius was fixed to one solar radius ($R_{\odot}$), and its limb darkening coefficients were fixed to the value of the Sun $u1=0.29 $ and $u2=0.34$ \citep{Claret-11}. The stellar effective temperature ($T_{\star}$) was fixed to the Sun value, 5778 K. The stellar inclination was fixed to 90 degree which means it is seen edge-on. Figure 3 shows the (near-)grazing transit light-curves of all four systems in the lines. In the next step we add a giant exactly polar spot (centered on the rotational axis) on the surface of the host star\footnote{Note that the exactly polar spot does not generate any out of transit modulation also, which may be mis-interpreted as an non-active star}. Then for each (near-)grazing planet we vary the filling factor and temperature of polar spots until the grazing transit light-curve vanish (as shown in the Figure 3 in the marked lines). Note that the criteria in order to consider a transit light-curve vanishes is that the transit's depth is smaller than 75 ppm \citep{Gilliland-11}. The spots' temperature contrast with the stellar temperature ($\Delta T= T_{\star}- T_{spot}$) required to vanish the grazing transit were estimated to be around 2500 K. The filling factor required for the planet on the lower impact parameters is bigger than for the higher impact parameter. For instance grazing planet with $b= 1.05$ required a polar spot with filling factor of $f=12\%$, and near-grazing planet with $b=0.90$ needed a polar spot with filling factor of $f=42\%$. It is also important to note that the highest spot's filling factor and temperature contrast required in our test, are smaller than the maximum observed polar spot's filling factor and temperature contrast. Therefore, our hypothesis does not required un-physical property for the polar spot to be valid.

\begin{figure}
\includegraphics[width=95mm, height=65mm, trim=10mm 0 0 0]{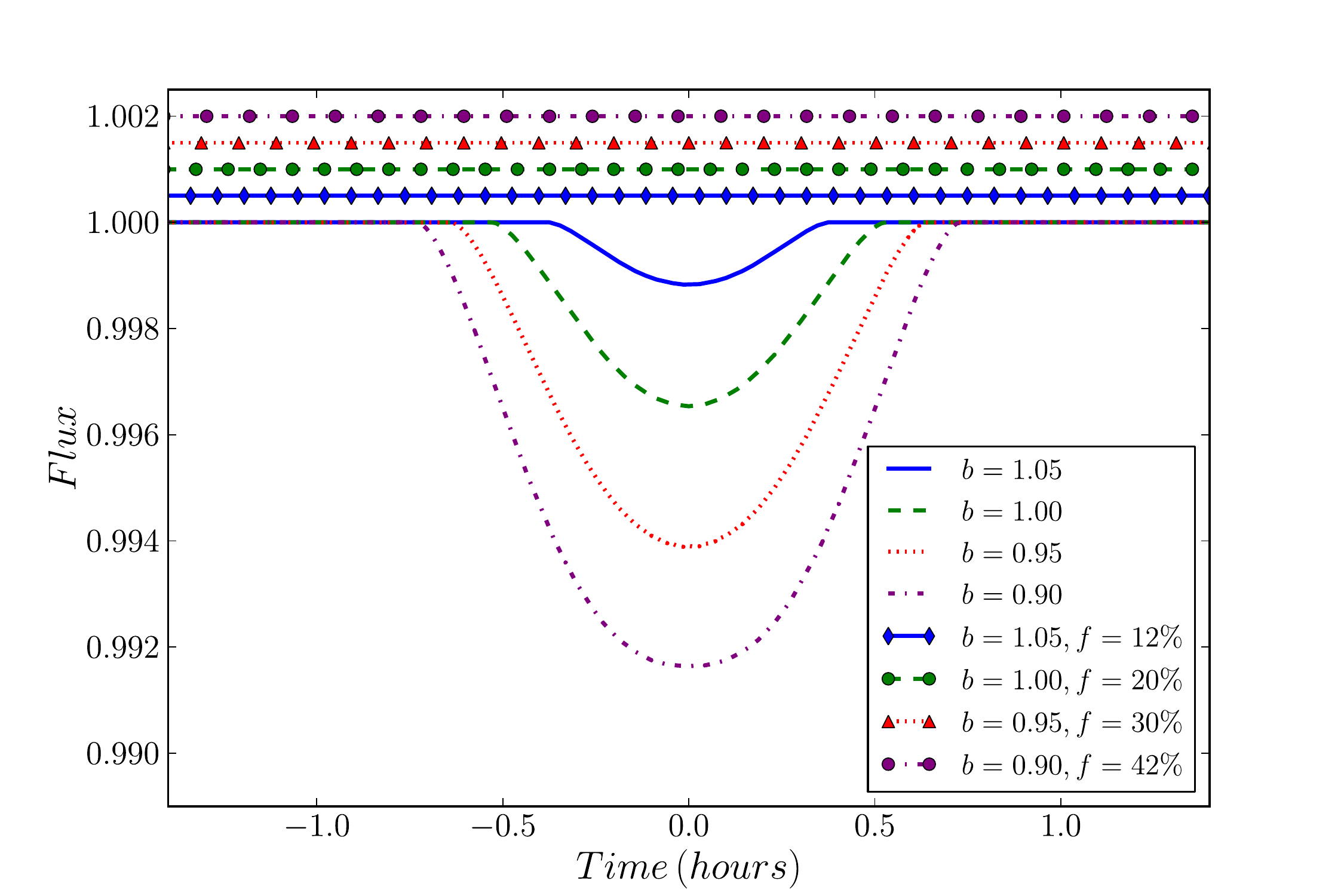}
 \caption{Disappearing the transit light-curve due to occultation of grazing transiting planet with the polar spots. The transit light-curves are generated with the SOAP-T tool. The lines show the grazing planet's transit light-curves correspond to different impact parameters($b=0.90,0.95,1.00, 1.05$) over a star without any spot. The line-markers show the grazing planets which occult a polar spots (centered on the stellar pole) with different filling factors ($f$). All the spots have temperature contrast with respect to the stellar temperature of $\Delta T=2500$ K. Note that the line-markers were vertically offset for clarity. }
  \label{sample-figure}
\end{figure}

\section{Conclusions}

In this paper, we assess a possible cause of low number of discovered (near-)grazing planetary transits. We compared the synthetic impact parameter distribution of detectable planets (through our simulation) with the observed impact parameter distribution. We found that a larger number of (near-)grazing should have been detected than what have been detected. Our hypothesis for the insufficient number of (near-)grazing planets is based on assumption that a large number of (near-)grazing planets transit the host stars with dark giant polar spot. As a consequence, the transit light-curves disappear due to the occultation of grazing planet and the polar spot. We would like to note that since the (near-)grazing transit light curves are shorter, shallower, and are often ``V-shape" (similar to the light-curve of a
eclipsing binary), they can therefore introduce an observational bias leading to a low number of detected (near-)grazing planets. Hereby, we would like to encourage transit hunter to perform more careful analysis on ``V-shape" transit candidates. 

Finally, we evaluated the filling factor of the required polar-spots, and we conclude that their filling factor are in a reasonable and physically feasible range.

\begin{acknowledgements}

\scriptsize MO acknowledges support by Centro de Astrof\' isica da Universidade do Porto through grant
with reference of
CAUP-15/2014-BDP. P.F., N.C.S., and S.C.C.B acknowledge the support from FCT through
Investigador FCT contracts of reference IF/01037/2013, IF/00169/2012, and IF/01312/2014, respectively, and POPH/FSE (EC) by FEDER funding through
the program “Programa Operacional de Factores de Competitividade -
COMPETE”. PF further acknowledges support from Funda\c{c}\~ao para a Ci\^encia e a Tecnologia (FCT) in the form of an exploratory project of reference
IF/01037/2013CP1191/CT0001. V.A. acknowledges the support from
the Funda\c{c}\~ao para a Ci\^encia e a Tecnologia (FCT) in the form of the ˆ
grant SFRH/BPD/70574/2010. AS is supported by the European Union under a Marie Curie Intra-European
Fellowship for Career Development with reference FP7-PEOPLE-2013-IEF,
number 627202. This work was supported by Funda\c{c}\~ao para a Ci\^encia e a Tecnologia (FCT) through the
research grant UID/FIS/04434/2013. We would like to thank the anonymous
referee for constructive comments and insightful suggestions.

\end{acknowledgements}

\bibliographystyle{aa}
\bibliography{mahlibspot}

\end{document}